\documentclass[preprint,12pt]{elsarticle}

\usepackage{amssymb}
\usepackage{amsthm}
\usepackage{amsmath}
\usepackage{algorithm}
\usepackage{algpseudocode}
\usepackage{booktabs}
\usepackage{multirow}
\usepackage{multicol}
\usepackage{booktabs}
\usepackage{color}
\usepackage{ccaption}
\usepackage{booktabs}
\usepackage{float}
\usepackage{fancyhdr}
\usepackage{caption}
\usepackage{xcolor,stfloats}
\usepackage{comment}
\usepackage{multirow}
\usepackage{booktabs}
\usepackage{array}
\usepackage{float}
\usepackage{ulem}
\usepackage{color}
\graphicspath{{figures/}}
\usepackage{cuted}
\usepackage{amsmath}
\usepackage{makecell}
\usepackage{array}
\usepackage{booktabs}

\journal{Nuclear Physics B}

\begin{document}

\begin{frontmatter}



\title{Enhancing Anti-spoofing Countermeasures Robustness through Joint Optimization and Transfer Learning}

\author[1,2]{Yikang Wang}
\author[2]{Xingming Wang}
\author[1]{Hiromitsu Nishizaki}
\author[2]{Ming Li\corref{cor1}}
\affiliation[1]{organization={Integrated Graduate School of Medicine, Engineering, and Agricultural Sciences, University of Yamanashi},
            addressline={4-4-37, Takeda},
            city={Kofu},
            postcode={400-8510},
            state={Yamanashi},
            country={Japan}}

\affiliation[2]{organization={Suzhou Municipal Key Laboratory of Multimodal Intelligent Systems, Duke Kunshan University},
            addressline={No. 8 Duke Avenue},
            city={Kunshan},
            postcode={215316},
            state={Jiangsu},
            country={China}}

\affiliation[3]{organization={School of Computer Science, Wuhan University},
            addressline={299 Bayi Road},
            city={Wuchang District, Wuhan},
            postcode={430072},
            state={Hubei},
            country={China}}

\cortext[cor1]{Corresponding author}

\begin{abstract}
Current research in synthesized speech detection primarily focuses on the generalization of detection systems to unknown spoofing methods of noise-free speech. However, the performance of anti-spoofing countermeasures (CM) system is often don't work as well in more challenging scenarios, such as those involving noise and reverberation. To address the problem of enhancing the robustness of CM systems, we propose a transfer learning-based speech enhancement front-end joint optimization (TL-SEJ) method, investigating its effectiveness in improving robustness against noise and reverberation. We evaluated the proposed method's performance through a series of comparative and ablation experiments. The experimental results show that, across different signal-to-noise ratio test conditions, the proposed TL-SEJ method improves recognition accuracy by 2.7\% to 15.8\% compared to the baseline. Compared to conventional data augmentation methods, our system achieves an accuracy improvement ranging from 0.7\% to 5.8\% in various noisy conditions and from 1.7\% to 2.8\% under different RT60 reverberation scenarios. These experiments demonstrate that the proposed method effectively enhances system robustness in noisy and reverberant conditions.
\end{abstract}

\begin{graphicalabstract}
\end{graphicalabstract}

\begin{highlights}
\item Research highlight 1
\item Research highlight 2
\end{highlights}


\end{frontmatter}


\section{Introduction}
\label{intro}
\begin{figure*}
  \centering
  \includegraphics[width=1\linewidth]{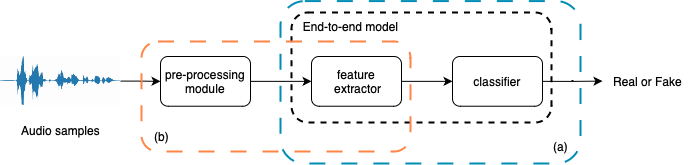}
  \caption{The illustration of typtical pipeline solution for synthesized speech detection systems. (a) The clean data-based spoofing detection systems are mainly implemented on these parts. (b) Noise data-based spoofing detection systems are mainly implemented on these parts}
  \label{fig:pipline}
\end{figure*}

With the rapid advancement of artificial intelligence, voice conversion (VC) and text-to-speech (TTS) technologies leveraging deep learning have achieved remarkable progress \cite{aguero2006, borsos2023, mehta2023, alexanderson2023, vovk2022, le2023, mehta2024}. These speech synthesis technologies can now produce high-quality, natural, and expressive human-like speech that is nearly indistinguishable from actual human voices. While these innovations provide significant benefits across various domains, their potential misuse poses serious threats to automatic speaker verification (ASV) systems, and can endanger social security, political stability, and economic integrity. As a result, there has been considerable focus on developing synthesized speech anti-spoofing countermeasure (CM) systems to address these risks. A typical CM system pipeline, as shown in Figure~\ref{fig:pipline}, includes a pre-processing module, a feature extractor, and a back-end classifier. If the feature extraction operation and the back-end classifier are integrated, with the waveform as the input, it can be considered an end-to-end model.

Since 2015, the ASVspoof community has organized anti-spoofing challenges biennially, holding four events consecutively \cite{asvspoof15, asvspoof17, asvspoof19, asvspoof21}. These challenges have significantly advanced spoofing detection technologies. Among them, the ASVspoof 2019 challenge is notable for its comprehensive approach, addressing all three major spoofing attack types: text-to-speech (TTS), voice conversion (VC), and replay attacks. The logical access (LA) dataset from ASVspoof 2019, widely used in current research, includes genuine speech from the Voice Cloning Toolkit (VCTK) corpus \cite{vctk}. The spoofed data is generated using advanced TTS and VC technologies developed in recent years, utilizing speech data from other speakers within the VCTK dataset to prevent data leakage. Since the VCTK corpus was recorded in a studio environment, the ASVspoof 2019 LA dataset is also widely regarded as a clean, noise-free dataset.


In studies based on clean datasets, as illustrated in Figure~\ref{fig:pipline}(a), various strategies have been explored at the feature extractor or back-end classifier stages to enhance the performance of CM systems for synthesized speech detection. For the feature extractor stage, the purpose of feature extraction is to learn discriminative features by capturing audio artifacts present in speech signals. Researchers have investigated the impact of different hand-crafted or prosodic features as inputs and have achieved moderate performance. Such as Constant Q Transform (CQT) \cite{}, Constant Q Cepstral Coefficients (CQCC) \cite{}, Linear Frequency Cepstral Coefficients (LFCC) \cite{}, Log Mel-Filter Bank (FBANK), and fundamental frequency (F0) \cite{}. Additionally, neural network-based front-ends have been utilized to extract deep features from raw waveform input. Tak et al. \cite{} employed SincNet \cite{} as the first layer of their end-to-end anti-spoofing model called RawNet2. Furthermore, self-supervised embedding features from pre-trained models like Wav2vec \cite{} and Hubert \cite{}, which are trained using only bona fide speech data, have been explored for their potential to enhance spoofing detection performance.


In the back-end classifier phase, diverse classification approaches have been explored, including conventional Gaussian Mixture Model (GMM) \cite{} and Neural Network (NN)-based models, such as Light Convolutional Neural Network (LCNN) \cite{}, Deep Residual Network (ResNet) \cite{}, and Graph neural networks (GNNs) \cite{}. GMM has been widely used as a baseline model in a series of competitions, such as ASVspoof 2017, 2019, and 2021. However, most NN-based models outperform the GMM-based classifier due to their powerful modeling capabilities \cite{}. In addition to improving CM system performance through model architecture enhancements, there are also approaches that modify training strategies or introduce new loss functions. For instance, Xue et al. \cite{} proposed a self-distillation network, where the deepest network instructs shallower networks to enhance their ability to capture fine-grained information. Wang et al. \cite{} introduced the adaptive and hyper-parameter-free Probability-to-Similarity Gradient (P2SGrad) into the synthesized speech detection task, utilizing the mean-square-error (MSE) loss function.

Moreover, some research has integrated feature extractor and classisifer into end-to-end systems. Notable examples of these comprehensive systems include Spectro-Temporal Graph Attention Network (RawGAT-ST) \cite{} , Audio Anti-Spoofing using Integrated Spectro-Temporal Graph Attention Networks (AASIST) \cite{} , and Rawformer \cite{}. These end-to-end approaches aim to modelling local and global artefacts and relationship directly on raw audio, enhance the perception of frequency and time domain features, and potentially improve overall detection performance.


Besides improving the detection precision of the CM system on clean data, some studies have also focused on enhancing the robustness of synthesized speech detection (SSD) tasks. For example, the ASVspoof 2021 logical access challenge was designed to explore SSD under audio channel coding and compression conditions. Research on robustness using this dataset mainly concentrates on the preprocessing and feature extraction phases of the pipeline. In the preprocessing phase, as shown in Figure~\ref{fig:method_comp}(a), data augmentation is the most commonly used technique to enhance model robustness. For instance, Tek et al. \cite{} proposed the RawBoost data augmentation method, which is tailored for telephony scenarios and requires no additional data sources, effectively improving the robustness of CM systems to compression. In the feature extraction phase, Wang et al. \cite{} introduced preprocessing methods such as band trimming, band-pass filtering, and band extension to identify suitable sub-band widths for coding and compression-robust SSD tasks. Furthermore, Tian et al. \cite{} generated their own training and test data subsets based on the ASVspoof 2015 dataset by adding noise and reverberation to the clean data, thereby exploring the robustness of CM systems with different feature inputs. Their investigations demonstrated that environmental noise and reverberation can significantly degrade the performance of CM systems.



This study builds upon our previous work \cite{}. In our earlier research, as shown in Figure~\ref{fig:method_comp}(b), we innovatively introduced speech enhancement as a front-end module to the audio anti-spoofing task, demonstrating its impact on the performance of back-end CM systems. Additionally, we proposed a joint training framework for the SSD task, optimizing both the Unet-based speech enhancement front-end and the audio anti-spoofing back-end models jointly, and conducted preliminary validations under specific noisy conditions. To the best of our knowledge, this is the first time that speech enhancement front-end have been employed for noise-robust SSD task. While the results were promising, the study had some limitations. Specifically, the previous research only addressed environmental noise and did not consider the impact of reverberation. Additionally, the data were generated offline for each noise condition, and separate models were trained for each scenario. This implementation setting limited the ability to assess the models' generalization performance and noise robustness in more realistic settings.


In this study, to improve performance in noisy scenarios, we first further investigate the impact of pre-trained models on both noise and reverberation robustness of CM systems. We propose a transfer learning-based speech enhancement front-end joint optimization (TL-SEJ) method. By employing transfer learning techniques at the back-end model, we integrate pre-trained information into the previously proposed joint training framework, as shown in Figure~\ref{fig:method_comp}(e), thereby enhancing the model's robustness to noise. Since self-supervised pre-trained models require raw audio as input and cannot be directly connected to the Unet front-end, we use the Conformer model pre-trained on the automatic speech recognition (ASR) task as the back-end model in this approach. This allows the CM system to retain the capability for joint optimization with speech enhancement while incorporating pre-trained information. Additionally, we use the Unet front-end's output noise mask instead of directly restoring spectral features, extending the system's robustness to non-additive noise such as reverberation. This approach aims to improve the overall performance and reliability of the CM system in various challenging acoustic environments. The key contributions of this work are as follows:

\begin{figure*}
  \centering
  \includegraphics[width=1\linewidth]{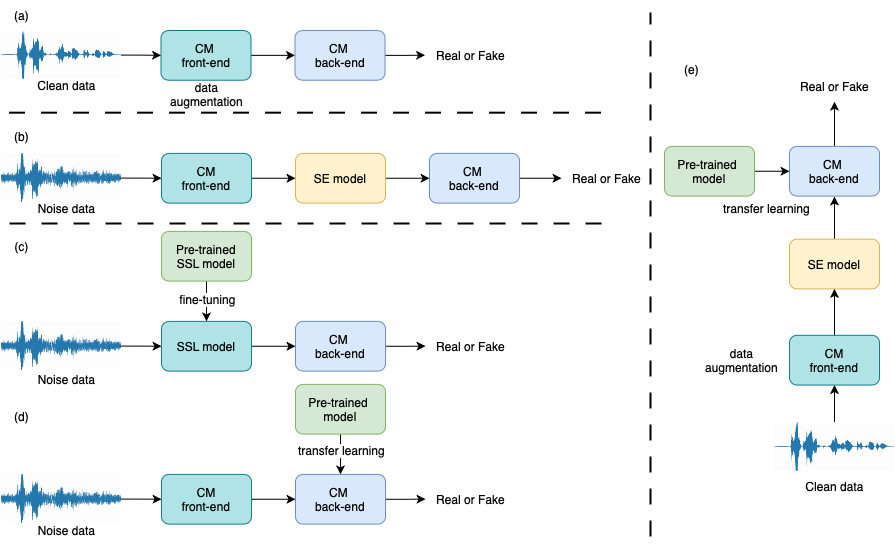}
  \caption{}
  \label{fig:method_comp}
\end{figure*}

\begin{itemize}
    \item We propose a transfer learning-based speech enhancement front-end joint optimization (TL-SEJ) method, enhancing the robustness of CM systems to both noise and reverberation.
    \item By integrating pre-trained information into the previously proposed joint training framework, we improve the model's robustness to noise using the Conformer model pre-trained on the ASR task as the back-end model.
    \item we propose a Dual-input Unet-based Masked feature Enhancement Network (DUMENet) as front-end SE module, where clean speech is concatenated with mixed speech as input, and masked features are used as output instead of directly reconstructing the original speech signal. This approach effectively handles non-additive noise, the reverberation, providing a robust enhancement for the CM system.
    \item We adopt a more comprehensive training approach, using unified models trained on mixed noise conditions. This enables us to evaluate the models' generalization performance and robustness more accurately. Additionally, we provide a comprehensive evaluation of the proposed TL-SEJ method, demonstrating its improved performance and reliability.
\end{itemize}

The remainder of this paper is organized as follows: Section II introduces the related work, Section III describes the proposed TL-SEJ method, Section IV introduce the experimental setup, and Section V presents and analysis the experimental results. Finally, Section VI concludes the paper. 

\section{Related Work}

To determine the authenticity of mixed speech signals with noise or reverberation, the most straightforward approach is to use data augmentation to enhance the diversity of training data, thereby improving the robustness of anti-spoofing networks, as shown in Figure 2(a). In our previous work, as illustrated in Figure 2(b), we introduced a speech enhancement front-end to the anti-spoofing task, achieving notable robustness in anti-spoofing CM systems. Additionally, as shown in Figure 2(c), models pre-trained on large datasets have learned prior knowledge from extensive and diverse data, which can significantly improve the robustness of downstream tasks. This section introduces related work in the fields of data augmentation, pre-trained models, and speech enhancement.

\subsection{Data Augmentation}

Data augmentation is a method used to generate additional training data for various neural networks. In speech-related tasks, references [9, 10] increased artificial data for low-resource ASR tasks. Reference [12] synthesized noisy audio by overlaying clean audio with noise signals. Reference [13] applied speed perturbation to the original Voxceleb audio to enrich the number of speakers. Reference [14] explored the use of an acoustic room simulator. In the SSD task, Qian et al. proposed multi-condition training based on data augmentation, allowing the model to access distorted training data. Cai and Li [15] used an online data augmentation method based on the MUSAN and RIR corpora to diversify frame-level data in deepfake boundary detection tasks. In this paper, for noise-based data augmentation, we follow the online noise simulation method using the MUSAN dataset as in [15]. However, for adding reverberation, we employ the latest online simulation methods to obtain richer room impulse responses (RIRs) and higher simulation efficiency.

\subsection{Pre-training Model and Transfer Learning}
In addition to SSL pre-trained models, supervised pre-trained models have also achieved notable results in many downstream tasks. For example Cai and Li \cite{} utilized a pre-trained ASR Conformer encoder \cite{} to initialize the speaker embedding network, thereby enhancing the model's generalization ability and reducing the risk of overfitting.
In this study, if we use an SSL pre-trained model as the backend following the speech enhancement network, it requires the speech enhancement network to output a complete time-domain audio signal. \cite{} indicates that fully reconstructing the time-domain signal may introduce new artifacts, increasing the training burden of the CM system. Moreover, SSL models usually have a large number of parameters, using an SSL model as a trainable backend rather than a freezing feature extractor requires fine-tuning, which significantly increases hardware and time costs. Therefore, we follow the transfer learning approach described in \cite{}, using Conformer encoder pre-trained on the ASR task. By aggregating different levels of feature embeddings through the multi-scale feature aggregation (MFA) method, we conduct supervised training of the Conformer model on the downstream SSD task. The specific transfer learning method is detailed in Section~\ref{par:approach}.
\subsection{Speehc Enhancement}

The aim of speech enhancement (SE) is to extract clean speech or to improve the speech-to-background ratio by removing noise and reverberation from a mixed speech signal \sout{cite{60year}}. For SE tasks, most studies focus on human hearing rather than machine hearing, where the time-domain waveform of the clean speech needs to be reconstructed \sout{cite{some mainstream SE papers}}. The problem of noise robustness cannot be ignored in any speech-related task, so many tasks use SE as an upstream process, using the reconstructed noise-free sound output from the SE network as input \cite{}. There are also tasks that integrate the SE network into the model and train it as a whole \cite{}, or enhance the model's speech enhancement capability by changing the loss function \cite{}. For the SSD task, Hanilci et al. \cite{} demonstrated that using noise-free sounds processed by a separate SE front-end as input to the CM system may increase the difficulty of training or degrade the system's performance. This performance degradation might be because speech enhancement introduces musical noise and other processing artifacts that mask the synthesis or conversion artifacts. Therefore, in our previous work, we introduced the SE module simultaneously with the CM system as part of a joint optimization process. Instead of performing a full time-domain waveform reconstruction, we computed the mean square error (MSE) at the spectral features (FBANK), which could avoid some reconstruction artifacts \cite{}.

\begin{figure*}
  \centering
  \includegraphics[width=1\linewidth]{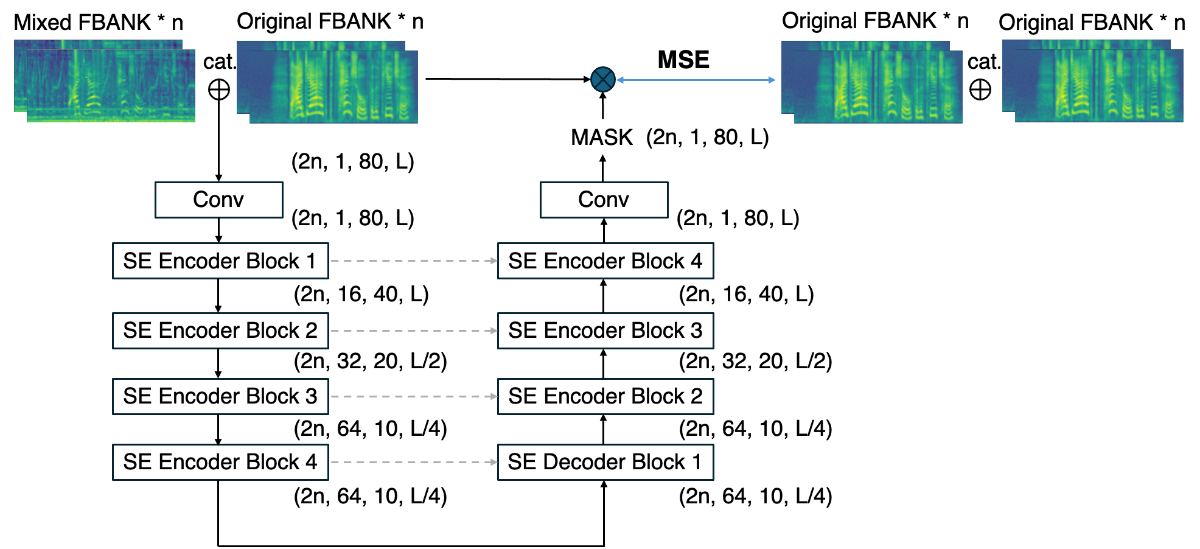}
  \caption{unet}
  \label{fig:unet}
\end{figure*}

U-Net is a classical model structure [11-15]. The basic components in this framework are the encoder, decoder, and skip connections. The encoder mainly down-samples the input speech features to obtain a higher-level speech embedding. The decoder is the inverse process of the encoder, up-sampling the speech feature map and ultimately outputting enhanced speech features with the same length as the original input. Skip connections link each encoding layer with its corresponding decoding layer, which helps to preserve features from each encoding layer to the decoding layer. Figure~\ref{fig:unet} illustrates the proposed Unet-based SE model used in this study. The specific model is detailed in Section~\ref{par:approach}.

\section{Methodology}
As shown in Figure~\ref{fig:method_comp}, to distinguish the authenticity of mixed speech signals with noise or reverberation, we summarize three training strategies that CM systems can use: data augmentation, speech enhancement, and utilizing pre-trained models. The methodology of this study is based on the SE front-end jointly optimisation method we previously proposed \cite{}. In this paper, we have proposed a Dual-input Unet-based Masked feature Enhancement Network (DUMENet) to address the reverberation better, and a transfer learning-based speech enhancement front-end joint optimization (TL-SEJ) mothod to introducing pre-training model back-end learned prior knowledge for large-sclae dataset. These enhancements aim to address the limitations observed in the initial study.
\subsection{Data Augmentation in the Preprocessing Stage}
\label{method:dataaug}
Data augmentation involves adding noise or reverberation to clean speech data, making the input speech more diverse. Models trained using data augmentation techniques have a stronger ability to handle complex signals. In this study, we use the officially provided ASVspoof 2019 LA train subset as the clean dataset, adding noise or reverberation to obtain the corresponding simulated mixed training data. A noisy speech signal can be represented as:
\begin{equation}
\label{eq:1}
x_n(t) = s(t) + n(t)
\end{equation}
where $x_n(t)$, $s(t)$, and $n(t)$ represent the mixed noisy speech, clean source speech, and noise signals, respectively. A reverberant speech signal $x_r(t)$ can be represented as:
\begin{equation}
x_r(t) = s^{(d)}(t) + s^{(e)}(t)
\end{equation}
It consists of the microphone directly received signal $s^{(d)}(t)$ and the late reverberation $s^{(e)}(t)$ generated after many reflections. According to \cite{}, to obtain simulated reverberant signals, we usually convolve the clean source speech with the room impulse response (RIR). Generally, the convolution operation can be expressed as:
\begin{equation}
\label{eq:3}
x_r(t) = s(t) * h(t)
\end{equation}
where $(*)$ denotes convolution calculation, $x_r(t)$ is the convolved reverberant speech signal, $s(t)$ is the original audio signal, and $h(t)$ is the RIR. The definition of convolution is:
\begin{equation}
   x_r(t) = \int_{-\infty}^{\infty} s(\tau) h(t - \tau) d\tau
\end{equation}
In the discrete case, the equation becomes:
\begin{equation}
   x_r[t] = \sum_{m=-\infty}^{\infty} s[m] h[n - m]
\end{equation}
When using data augmentation methods, we perform online noise and reverberation simulation on the data while reading it, making the training samples more diverse. We ensure that the noise samples used for the training and test sets are isolated to prevent data leakage. For details on the implementation of online data augmentation and the generation of test datasets, see Chapter~\ref{dataset}.

\subsection{Unet-based Speech Enhancement Before Anti-Spoofing Network}

The purpose of SE is to remove noise and reverberation from the mixed speech signal $x(t)$ and estimate the target clean speech $s(t)$. Using the basic Unet as an example, a traditional SE network applies the Short-Time Fourier Transform (STFT) to the input $x(t)$, decomposing the complex-valued spectrogram into magnitude and phase components. Only the magnitude is then fed into the Unet enhancement network, which returns an estimated magnitude spectrogram of the clean speech. To generate the corresponding audio signal, these magnitudes are combined with the mixed phase and converted back to the time-domain waveform $s_i(t)$ via inverse STFT or some hand-crafted vocoder algorithms. This process can be represented as follows:

\begin{equation}
|X(t, f)| e^{j \phi(t, f)} = \text{STFT}(x(t))
\end{equation}

\begin{equation}
|S_i(t, f)| = \text{Unet}(|X(t, f)|)
\end{equation}

\begin{equation}
s_i(t) = \text{iSTFT}(|S_i(t, f)| e^{j \phi(t, f)})
\end{equation}

where $Unet(\cdot)$ is the speech enhancement network, $|X(t, f)|$ is the magnitude spectrogram, and $\phi(t, f)$ is the phase spectrogram. For a pair of clean and mixed magnitude spectrograms $|S(t, f)|$ and $|X(t, f)|$, only the magnitude $|X(t, f)|$ is input into the speech enhancement model, and the Unet model returns the estimated magnitude spectrogram $|S_i(t, f)|$, where $f$ represents the frequency bin index and $t$ represents the time frame. The loss function $L_{mse}$ is designed to minimize the mean square error (MSE) between the clean spectrum and the recovered spectrum,

\begin{equation}
L_{mse} = \frac{1}{t \times f} \sum_t \sum_f ||~|S_i(t, f)| - |S(t, f)|~||_2^2 
\end{equation}
where $||\cdot||_2$ denotes the $L2$ norm.

In this paper, we propose DUMENet, as shown in Figure~\ref{fig:unet}, whose model structure is almost same as the Unet-based front-end SE network structure in our previous work \cite{}. Inspired by the work of xxx et al. \cite{} on the joint training of SE front-ends with speaker embedding extractors in the ASV task, we concatenates the FBANK features of noisy speech with clean speech as the input at every training minibatch. Additionally, it uses the FBANK features of the clean source speech, concatenated twice, as labels. This dual-feature input approach enhances the similarity between the extracted clean and noisy speech embedding features during joint optimization of the front-end and back-end, thus avoiding additional artefacts while suppressing noise perturbations. 
Unlike our previous work, DUMENet does not directly reconstruct the FBANK features of clean speech. Instead, it outputs a soft mask of the same shape as the input. This soft mask is element-wise multiplied with the input signal, and the resulting product is compared to the label using the MSE loss function. This approach aligns with the reverberation simulation principles mentioned in Section~\ref{method:dataaug}, further enhancing the performance of the CM system under reverberant conditions. The loss function $L_{mse'}$ for DUMENet can be expressed as:

\begin{align}
L_{mse'} = \frac{1}{t \times f} \sum_{t} \sum_{f} (&\| X_b(t, d)\cdot\textbf{Mask}_x(t, d)) - S_b(t, d) \|_2^2 \notag \\
+ &\| S_b(t, d)\cdot\textbf{Mask}_s(t, d)) - S_b(t, d) \|_2^2) 
\end{align}

where $X_b(t, d)$ and $S_b(t, d)$ represent the noisy and clean FBANK features with a time sequence length of $t$ and a Mel-scale filter bank dimension of $d$, respectively. $\textbf{Mask}_x$ and $\textbf{Mask}_s$ denote the soft masks output by DUMENet for noisy and clean inputs, respectively.

\subsection{Transfer Learning in the Anti-Spoofing Network Phase}
\label{par:approach}
\subsubsection{Conformer Model}
In the Conformer network, the input audio signals are processed through a feature extractor and a subsampling convolution layer to reduce the input length, then fed into the Conformer encoder composed of multiple Conformer blocks. Figure 1 presents the Conformer encoder and block structure. A Conformer block is a sandwich structure of four modules stacked together: a feed-forward (FFN) module, a multi-headed self-attention (MHSA) module, a convolution module, and a second FFN module.

The MHSA employs a relative sinusoidal positional encoding scheme from Transformer XL \sout{cite{dai2019transformer}} to improve generalization across varying input lengths. The subsequent convolution module contains a point-wise convolution and a gated linear unit (GLU) activation layer, followed by a single 1-D depth-wise convolution layer with batch normalization to train deeper models. Two half-step FFN layers replace the original FFN layer in the Transformer block, one before the attention layer and one after \sout{cite{gulati2020conformer}}. Mathematically, this means for a given input feature $\mathbf{g}_{i} \in \mathbb{R}^{d \times T}$ with dimension $d$ and time series length $T$, the output $\mathbf{h}_i$ of the $i$-th Conformer block is given by:

\begin{align}
    \tilde{g_{i}} & =g_{i}+\frac{1}{2} \operatorname{FFN}\left(g_{i}\right) \\
    g_{i}^{\prime} & =\tilde{g_{i}}+\operatorname{MHSA}\left(\tilde{g_{i}}\right) \\
    g_{i}^{\prime \prime} & =g_{i}^{\prime \prime}+\operatorname{Conv}\left(g_{i}^{\prime}\right) \\
    h_{i} & =\operatorname{Layernorm}\left(g_{i}^{\prime \prime}+\frac{1}{2} \operatorname{FFN}\left(g_{i}^{\prime \prime}\right)\right)
\end{align}

\subsubsection{Transfer Learning with ASR Pre-trained Conformer Model}

Transfer learning can effectively improve training efficiency and introduce additional data information, enhancing model robustness.

Unlike Wang Xing et al., who used the large-scale self-supervised pre-trained model wav2vec2.0 as a front-end feature extractor for fine-tuning, performing transfer learning directly on the audio anti-spoofing back-end model significantly alters the distribution of the pre-trained model's output representations, enabling more direct domain matching for complex audio anti-spoofing tasks.

In this study, we use the Conformer model trained on ASR tasks as the pre-trained model for the audio anti-spoofing network back-end. The Conformer model has achieved excellent results in ASR tasks due to its integration of CNN modules into the Transformer encoder, making it proficient at modeling both local and global dependencies in speech signals.

We employ the MFA-Conformer framework proposed by Yang et al. as the back-end of the audio anti-spoofing network. As shown in Figure 1, this framework concatenates the output feature maps of all $L$ Conformer blocks to form a large feature map $\mathbf{H}'$:
\begin{equation}
    \mathbf{H}' = \text{Concat}(\mathbf{h}_1, \mathbf{h}_2, \ldots, \mathbf{h}_L)
\end{equation}
where $\mathbf{H}'$ has dimensions $\mathbb{R}^{D \times T}$, with $D = d \times L$.
The concatenated feature map $\mathbf{H}'$ is then normalized using LayerNorm to obtain $\mathbf{H}$:
\begin{equation}
  \mathbf{H} = \text{LayerNorm}(\mathbf{H'})  
\end{equation}
Specifically, for each time step $t$, the feature $\mathbf{H'}_t \in \mathbb{R}^D$ is normalized by its mean and variance. The LayerNormed $\mathbf{H}$ retains the original shape $\mathbb{R}^{D \times T}$, but with normalized feature values.
The next step is to apply attentive statistics pooling and a fully connected layer to obtain utterance-level spoofing detection embeddings. For the frame-level feature $\mathbf{H}_t$ at each time step $t$, the attention scores and normalized scores are first computed, and then these scores are used to calculate the weighted mean vector $\tilde{\mu}$ and weighted standard deviation $\tilde{\sigma}$.

\begin{algorithm}
\caption{Transfer Learning Algorithm}
\begin{algorithmic}[1]
\State \textbf{Input:} Pre-trained Conformer encoder parameters $\theta_{ASR}$, audio anti-spoofing dataset $\mathcal{D}_{AS}$
\State \textbf{Output:} Trained anti-spoofing CM backend $\theta_{CM}$ and linear classifier $f(W, b)$

\State Load $\theta_{ASR}$ into the Conformer encoder $\theta_{C}$ with the same network structure
\For {each Conformer block $i$}
    \State Compute output $h_i = \text{ConformerBlock}(g_i)$
\EndFor
\State Concatenate outputs: $H' = [h_1, h_2, \ldots, h_n]$
\State Apply LayerNorm: $H = \text{LayerNorm}(H')$
\State Apply Attentive Statistics Pooling (ASP):
\[
\alpha_t = \frac{\exp(e_t)}{\sum_{\tau=1}^T \exp(e_\tau)}, \quad \text{where } e_t = \mathbf{v}^T f(\mathbf{W}\mathbf{H}_t + \mathbf{b}) + k
\]
\State Compute weighted mean and standard deviation:
\[
\tilde{\mu} = \sum_{t=1}^T \alpha_t \mathbf{H}_t
\]
\[
\tilde{\sigma} = \sqrt{\sum_{t=1}^T \alpha_t \mathbf{H}_t \odot \mathbf{H}_t - \tilde{\mu} \odot \tilde{\mu}}
\]
\State Concatenate $\tilde{\mu}$ and $\tilde{\sigma}$ to form $H$
\State Attach a fully connected (FC) layer to extract high-level representations: $H = \text{FC}(H)$
\State Attach a linear classifier: $y = f(W \cdot H + b)$
\State Fine-tune $\theta_{CM}$ and classifier $f(W, b)$ using binary cross-entropy loss

\end{algorithmic}
\end{algorithm}

The transfer learning process is shown in Algorithm 1. We load the pre-trained Conformer encoder parameters from the ASR task into the corresponding Conformer encoder of the MFA-Conformer structure for the audio anti-spoofing network back-end. Using multi-scale feature aggregation (MFA), we concatenate the output feature maps from each Conformer block, apply layer normalization to obtain frame-level high-level representations, and input these representations into an attentive statistics pooling (ASP) layer and a fully connected (FC) layer to obtain utterance-level spoofing detection embeddings. Finally, a linear classifier is attached for fine-tuning. Similar to training a general audio anti-spoofing network, we fine-tune the Conformer encoder and the linear classifier using a binary cross-entropy loss function on the audio anti-spoofing dataset, as given by:

\begin{equation}
L_{BCE} = \sum_i -(y_i \log(p_i) + (1 - y_i) \log(1 - p_i))
\end{equation}

where $ y_i \in \{0, 1\} $ represents the class labels and $ p_i $ represents the classifier's probability output.

\subsection{Joint Training of Speech Enhancement Front-End and Anti-Spoofing Back-End Models}
\subsubsection{Audio Anti-Spoofing Module}
To compare the effectiveness of joint optimization, in addition to the MFA-Conformer mentioned in the previous section, we also use two different network structures as the audio anti-spoofing back-end models: LCNN [24] and ResNet18 [25]. Both structures are commonly used in the ASVSpoof 2021 challenge for audio anti-spoofing. The Max-Feature-Map (MFM) operation, based on the Max-Out activation function, is a fundamental component of LCNN. A Bi-LSTM layer is used in LCNN to aggregate utterance-level embeddings. For ResNet18, it is a lightweight version of ResNet. Detailed model structures and parameters of LCNN and ResNet can be found in Chapter~\ref{exp}. Similarly, to ensure fair comparison, all audio anti-spoofing back-ends in this study use the Attentive Statistic Pooling (ASP) [14] block to aggregate utterance-level embeddings. Binary cross-entropy loss, as shown in Equation 12, is used as the objective function.

\subsubsection{Joint Training with Transfer Learning-Based Back-End}
As shown in Figure~\ref{fig:method_comp}(e), the TL-SEJ combines the three previously mentioned schemes by using on-the-fly data augmentation at the pre-processing stage to increase the diversity of input speech. The audio anti-spoofing back-end network uses a pre-trained Conformer model based on the ASR task to improve generalization performance. In between, we incorporate DUMENet to jointly optimize with the back-end network during training. This aims to reduce unknown additional artifacts due to the inconsistency between independent speech enhancement and anti-spoofing tasks, making the entire CM system more conducive to the deep embedding of the SSD task. Similar to the joint optimization loss used in our previous work, TL-SEJ also uses MSE loss and CE loss as a combined loss function, as shown below:

\begin{equation}
L = L_{ce} + L_{mse}
\end{equation}

\section{Dataset}
\label{dataset}
The detailed statistics of the databases used in this research are outlined in Table 3. The ASVspoof database is a series of data from the ASVspoof challenges \cite{asvspoof15,asvspoof17,asvspoof19,asvspoof21}. Among them, 19LA is one of the most widely used datasets by industry researchers. For this research, we use the ASVspoof 2019 LA (19LA) \cite{asvspoof19} datasets as our primary experimental dataset. Noisy and reverberant datasets are generated based on this dataset. The 19LA dataset consists mostly of clean data created using utterances from 107 speakers (46 male, 61 female) from the VCTK dataset \cite{}. These 107 speakers are partitioned into three speaker-disjoint sets for training, development, and evaluation. The spoofed utterances were generated using four TTS and two VC algorithms in the training and development sets, while 13 TTS/VC algorithms are used in the evaluation set, 4 of which are partially known and 7 of which are unknown for training and development.

\subsection{On-the-Fly Data Augmentation During Training}
In this study, two different on-the-fly data simulation strategies were used for noise and reverberation data augmentation during training to ensure diversity in the training samples while preventing data leakage.

1) Additive Noise Augmentation: We selected the MUSAN dataset [56] as our noise source, which contains approximately 60 hours of English real speech, 42 hours and 31 minutes of various music styles, and about 6 hours of various machine and environmental noises. We selected only a few hundred samples from each noise category as the source noise for training. As shown in Equation \ref{eq:1}, we add environmental noise, music, and babble noise $n(t)$ to the clean audio $s(t)$ to obtain the noisy speech $x(t)$. Babble noise is generated by combining three to eight independent files from the Speech subset $S(t)$. This process can be expressed as:
\begin{equation}
    x_{babble}(t) = s(t) + n_{babble}(t) = s(t) + \sum_{i=1}^{k} S_i(t), \quad k \in [3, 8]
\end{equation}

where $n_{\text{babble}}(t)$ represents the combined babble noise, and $\forall S_i(t) \in S(t)$. During the noise addition process, only one type of noise is used each time, with each noise type having an equal probability of 1/3 for data augmentation. The Signal-to-Noise Ratio (SNR) is randomly chosen from a range of 0 to 20 dB.

2) Convolutional Reverberation Noise Augmentation: According to Equation~\ref{eq:3}, we can obtain reverberant speech by convolving the clean speech $s(t)$ with the room impulse response (RIR) $h(t)$. Luo et al. proposed an RIR simulation tool called FRAM-RIR, which efficiently performs online reverberation augmentation on clean audio. When using this tool, we modified two parameters: we set the rectangular room dimensions to $3m \times 3m \times 2.5m \sim 10m \times 6m \times 4m$, and the reverberation time (RT60) was randomly chosen between $0.2s \sim 1.0s$. RT60 is the main room acoustic parameter, representing the time required for the sound energy in a room to decrease by 60 dB after the source emission has stopped. Generally, a larger RT60 indicates stronger reverberation.

To maintain variability during the training cycle, we integrated on-the-fly data augmentation, applying the above noise augmentation to each training speech sample with a probability of 0.7, meaning 70

\subsection{Preparation of Test Datasets}
To distinguish the effects of noisy and reverberant data on audio anti-spoofing detection, we created two test datasets by adding noise and reverberation to the ASVspoof 19LA evaluation set offline.

1) Noise Test Datasets: 
To create noise conditional evaluation sets, we artificially sample the last MUSAN noise signals from each category and add them to clean audio at 5 different SNRs: 0dB, 5dB, 10dB, 15dB, and 20dB. This results in a total of $3 \times 5 = 15$ noise test datasets, as shown in Table 3.

2) Reverberation Test Datasets:
Although the study by Tom et al. [57] provides about 40,000 simulated RIRs, they roughly categorize rooms into large, medium, and small sizes. To compare the effects under different reverberation conditions, we use RT60 as an indicator of reverberation intensity. When creating reverberation test datasets, we used the pyroomacoustics toolkit to generate RIRs offline. During the simulation, we randomly set the room dimensions to $10m \times 8m \times 2.8m \sim 15m \times 10m \times 4m$, and used 4 different RT60s: 0.25s, 0.5s, 0.75s, and 1s. This resulted in 4 reverberation test datasets, as shown in Table 3.

\subsection{Additional Test Dataset}
Additionally, we used the ASVspoof 2021 LA evaluation subset to test the out-of-domain performance of the proposed method. The 21LA evaluation set consists of data from various telephone transmission systems, including Voice over Internet Protocol (VoIP) and the Public Switched Telephone Networks (PSTN), thus exhibiting real-world signal and transmission channel effects. We use this dataset to observe the performance changes of the proposed method when encountering non-additive noise interferences such as channel coding.

\section{Experiments}
\label{exp}
\subsection{Experimental Setup}
The experimental setup of this study is as follows:
\begin{itemize}
    \item First, we aim to verify that the introduction of pre-trained models alone can improve system robustness. We test whether the spoofing detection systems with the two pre-training modes shown in Figure 1 perform better on noisy and reverberant test data compared to systems without pre-training.
    \item Second, we compare the LCNN, ResNet, and pre-trained Conformer backbone networks under clean training, data augmentation, and joint optimization with speech enhancement conditions to verify the performance improvement brought by joint optimization for noisy and reverberant test data.
    \item Third, we conduct ablation experiments on the Conformer pre-trained backbone network with joint optimization of speech enhancement.
    \item Finally, we test the out-of-sample performance of the models on the ASVspoof21LA evaluation dataset and the FAD noise unseen test dataset and compare them with baseline models.
\end{itemize}

\subsection{Experimental Configuration}
\subsubsection{Model Parameters and Training Conditions}

In the first part of the experiments, we used the end-to-end solution AASIST proposed by Jung et al. (2022) as the baseline model for the pre-training method shown in Figure 1(a). This model uses a novel heterogeneous stacking graph attention layer that models artifacts spanning heterogeneous temporal and spectral domains with a heterogeneous attention mechanism and a stack node. This method achieved the best EER and the lowest t-DCF in the ASVspoof19-LA competition. In contrast, the pre-training method proposed by Tek et al. (2021) uses the pre-trained wav2vec 2.0 XLS-R model as the front-end, replacing the sinc-layer front-end in AASIST. We refer to this method as SSL-AASIST, which obtains the state-of-the-art results reported in the literature for the ASVspoof 2021 LA database. For the pre-training model transfer learning method shown in Figure 1(b), we use the NEMO Conformer-CTC-small-ls model version 1.0.0\footnote{https://catalog.ngc.nvidia.com/orgs/nvidia/teams/nemo/models/stt\_en\_conformer\_ctc\_small\_ls} as the pre-trained CM backend model. This model has the same structure as [24] but replaces the Conformer transducer in [24] with a linear decoder backend and links a connectionist temporal classification (CTC) for decoding. According to the open source code, the convolutional layer of the Conformer-CTC-small-ls model has a downsampling rate of 1/4. The encoder part has 13M parameters with 4 attention heads and 16 Conformer blocks. The feature dimension of the encoder convolutional layer is 176, and the feature dimension of the FFN is 704. The model was trained on the LibriSpeech dataset, which contains around 1000 hours of English speech. [26, 33] We use this non-pre-trained Conformer model as a comparative baseline model.

As illustrated in Table 4, in the second part of the experiments, the LCNN and ResNet18 models are implemented, with the embedding size set to 256. The Squeeze-and-Excitation block [27] is also used in the ResNet18 model, and the dimension of the bottleneck in the Squeeze-and-Excitation block is set to 256. A more detailed architecture description can be found in Table 4.

\begin{table}[tb]
    \footnotesize
    \caption{The architecture of ResNet18, $\mathbf{C}$ denotes the convolutional layer, $\mathbf{S}$ denotes the shortcut convolutional layer.}
    \label{tab:res18}
    \centering
    \begin{tabular}[c]{@{\ \ \ }l@{\ \ \ }c@{\ \ \ }l@{\ \ \ }}
        \toprule
        \textbf{Layer} & \textbf{Output} & \textbf{Structure(kernal size, stride)} \\
        \midrule
        Conv1 & $16 \times D \times L$ & $\mathbf{C}(3\times 3, 1)$ \\
        \midrule
        \makecell[l]{Residual\\Layer 1} & $16 \times D \times L$ & $\begin{bmatrix}
            \mathbf{C}(3\times 3, 1) \\
            \mathbf{C}(3\times 3, 1)
        \end{bmatrix} \times 2$ \\
        \midrule
        \makecell[l]{Residual\\Layer 2} & $32 \times \frac{D}{2} \times \frac{L}{2}$ & $\begin{bmatrix}
            \mathbf{C}(3\times 3, 2) \\
            \mathbf{C}(3\times 3, 1) \\
            \mathbf{S}(1\times 1, 2)
        \end{bmatrix} \begin{bmatrix}
            \mathbf{C}(3\times 3, 1) \\
            \mathbf{C}(3\times 3, 1)
        \end{bmatrix}\times 2$ \\
        \midrule
        \makecell[l]{Residual\\Layer 3} & $64 \times \frac{D}{4} \times \frac{L}{4}$ & $\begin{bmatrix}
            \mathbf{C}(3\times 3, 2) \\
            \mathbf{C}(3\times 3, 1) \\
            \mathbf{S}(1\times 1, 2)
        \end{bmatrix} \begin{bmatrix}
            \mathbf{C}(3\times 3, 1) \\
            \mathbf{C}(3\times 3, 1)
        \end{bmatrix}\times 2$ \\
        \midrule
        \makecell[l]{Residual\\Layer 4} & $128 \times \frac{D}{8} \times \frac{L}{8}$ & $\begin{bmatrix}
            \mathbf{C}(3\times 3, 2) \\
            \mathbf{C}(3\times 3, 1) \\
            \mathbf{S}(1\times 1, 2)
        \end{bmatrix} \begin{bmatrix}
            \mathbf{C}(3\times 3, 1) \\
            \mathbf{C}(3\times 3, 1)
        \end{bmatrix}\times 2$ \\
        \midrule
        Pooling & $128 \times \frac{D}{8}$ & Attentive Statistics Pooling \\
        Linear & $128$ & Fully Connected Layer\\
        Linear & $2$ & Fully Connected Layer\\
        \bottomrule
    \end{tabular}
\vspace{-1em}
\end{table}

\begin{table}[tb]
    \footnotesize
    \centering
    \caption{The architecture of LCNN .\vspace{-0.6em}}
    \label{tab:lcnn}
    \setlength{\tabcolsep}{+1.4mm}{\begin{tabular}{lll} 
    \toprule
    \textbf{Layer} & \textbf{Structure(kernal size/stride)} & \textbf{Output Size} \\ 
    \midrule
    Conv\_1 & $5 \times 5 / 1 \times 1$ & $D \times L \times 64$ \\
    MFM\_2 & $-$ & $D \times L \times 32$ \\ \midrule
    MaxPool\_3 & $2 \times 2 / 2 \times 2$ & $D/2 \times L/2 \times 32$ \\ \midrule
    Conv\_4 & $1 \times 1 / 1 \times 1$ & $D/2 \times L/2 \times 64$ \\
    MFM\_5 & $-$ & $D/2 \times L/2 \times 32$ \\
    BatchNorm\_6 & $-$ & $D/2 \times L/2 \times 32$ \\
    Conv\_7 & $3 \times 3 / 1 \times 1$ & $D/2 \times L/2 \times 96$ \\
    MFM\_8 & $-$ & $D/2 \times L/2 \times 48$ \\ \midrule
    MaxPool\_9 & $2 \times 2 / 2 \times 2$ & $D/4 \times L/4 \times 48$ \\
    BatchNorm\_10 & $-$ & $D/4 \times L/4 \times 48$ \\ \midrule
    Conv\_11 & $1 \times 1 / 1 \times 1$ & $D/4 \times L/4 \times 96$ \\
    MFM\_12 & $-$ & $D/4 \times L/4 \times 48$ \\
    BatchNorm\_13 & $-$ & $D/4 \times L/4 \times 48$ \\
    Conv\_14 & $3 \times 3 / 1 \times 1$ & $D/4 \times L/4 \times 128$ \\
    MFM\_15 & $-$ & $D/4 \times L/4 \times 64$ \\ \midrule
    MaxPool\_16 & $2 \times 2 / 2 \times 2$ & $D/8 \times L/8 \times 64$ \\ \midrule
    Conv\_17 & $1 \times 1 / 1 \times 1$ & $D/8 \times L/8 \times 128$ \\
    MFM\_18 & $-$ & $D/8 \times L/8 \times 64$ \\
    BatchNorm\_19 & $-$ & $D/8 \times L/8 \times 64$ \\
    Conv\_20 & $3 \times 3 / 1 \times 1$ & $D/8 \times L/8 \times 64$ \\
    MFM\_21 & $-$ & $D/8 \times L/8 \times 32$ \\
    BatchNorm\_22 & $-$ & $D/8 \times L/8 \times 32$ \\
    Conv\_23 & $1 \times 1 / 1 \times 1$ & $D/8 \times L/8 \times 64$ \\
    MFM\_24 & $-$ & $D/8 \times L/8 \times 32$ \\
    BatchNorm\_25 & $-$ & $D/8 \times L/8 \times 32$ \\
    Conv\_26 & $3 \times 3 / 1 \times 1$ & $D/8 \times L/8 \times 64$ \\
    MFM\_27 & $-$ & $D/8 \times L/8 \times 32$ \\ \midrule
    MaxPool\_28 & $2 \times 2 / 2 \times 2$ & $D/16 \times L/16 \times 32$ \\
    \midrule
    Pooling & BiLSTM & $80+80$ \\
    Pooling & Attentive Statistics Pooling & $320$ \\
    Linear  & Fully Connected Layer & $128$ \\
    Linear  & Fully Connected Layer & $2$ \\
    \bottomrule
    \end{tabular}}
\vspace{-1em}
\end{table}

Table 5 shows the detailed architecture of the U-net based SE model. The total number of blocks is set to 8, with 4 blocks in the encoder and 4 blocks in the decoder. The number of channels for each layer in the encoder is set to 16, 32, 64, and 128. Additionally, the U-Net has one convolution layer and one transposed convolution layer.

For feature extraction, the logarithmic Mel-spectrogram is extracted by applying 80 Mel filters on the spectrogram computed over Hamming windows of 64ms shifted by 8ms. For joint training, the learning rate is set to 1e-3 during training. We adopt the ReduceOnPlateau learning rate (LR) scheduler with an initial LR of 0.1. All models are trained using the Adam optimizer.

\subsection{Evaluation Metrics}

According to the ASVspoof 2019 evaluation plan \sout{cite{EER2}}, \sout{cite{EER3}}, the evaluation metric for the countermeasures in this study is the equal error rate (EER). Let $P_{fa}(\theta)$ and $P_{miss}(\theta)$ denote the false alarm and miss rates at threshold $\theta$:

\begin{equation}
p_{fa}(\theta_{EER}) = p_{miss}(\theta_{EER}) 
\end{equation}

\begin{equation}
P_{fa}(\theta) = \frac{\#\{\text{fake trials with score} > \theta\}}{\#\{\text{total fake trials}\}}
\end{equation}

\begin{equation}
P_{miss}(\theta) = \frac{\#\{\text{real trials with score} \leq \theta\}}{\#\{\text{total real trials}\}}
\end{equation}

EER corresponds to the threshold $\theta_{EER}$ at which the two detection error rates are equal, i.e., $EER = P_{fa}(\theta_{EER}) = P_{miss}(\theta_{EER})$. The lower the value of EER, the better the performance of the model.

\subsection{Experimental Results}
\subsubsection{Comparison of Pre-trained Models with Clean Training}
We first verify the robustness and limitations of pre-trained models under noise and reverberation conditions. As shown in Table 6, under the condition of using only the clean training data provided by ASVspoof 2019 LA, the Conformer-based anti-spoofing system with our proposed pre-trained transfer learning method and the front-end based AASIST anti-spoofing system with large-scale pre-trained models demonstrate better noise robustness compared to their respective baseline models. However, under reverberation conditions, the transfer learning scheme shows more significant improvement over the SOTA model's pre-trained fine-tuning scheme only under low reverberation conditions with RT60 of 0.25s. As the reverberation increases, the transfer learning scheme significantly reduces the reverberation robustness of the anti-spoofing system. This is because the LibriSpeech dataset used for pre-training the Conformer contains no noise or reverberation, whereas the XLSR front-end is pre-trained on a larger dataset that includes more complex data scenarios. This is also reflected in the Music and Environmental Noise test sets, where the performance improvement of the Conformer pre-trained model is less significant than that of the SSL-AASIST scheme. Therefore, although pre-training exposes the model to more diverse data, pre-training alone is not sufficient to provide the model with enough generalization ability to handle various types of disturbances.

\begin{table}[tb]
\footnotesize
\centering
\caption{\%EER of }
\label{tab:4 SSL}
\begin{tabular}{llllll}
\toprule \toprule
\multicolumn{2}{l}{\textbf{model}}                    & \multicolumn{1}{c}{\textbf{Conformer}} & \multicolumn{1}{c}{\textbf{T. Conformer}} & \multicolumn{1}{c}{\textbf{AASIST}} & \multicolumn{1}{c}{\textbf{SSL-AASIST}} \\ \hline \hline
\multirow{19}{*}{test dataset} & Babble 20dB & 13.5                          & \textbf{5.48}                    & 7.97                       & \textbf{5.67}                  \\
                               & Babble 15dB & 16.04                         & \textbf{9.21}                    & 12.21                      & \textbf{10.28}                 \\
                               & Babble 10dB & 19.82                         & \textbf{15.24}                   & 23.04                      & \textbf{17.23}                 \\
                               & Babble 5dB  & 25.13                         & \textbf{24.35}                   & 36.04                      & \textbf{25.33}                 \\
                               & Babble 0dB  & 29.81                         & \textbf{34.19}                   & 44.74                      & \textbf{29.8}                  \\ \cmidrule(lr){2-2} \cmidrule(lr){3-4} \cmidrule(lr){5-6}
                               & Music 20dB  & 17.2                          & \textbf{4.58}                    & 7.44                       & \textbf{1.03}                  \\
                               & Music 15dB  & 22.04                         & \textbf{6.48}                    & 13.35                      & \textbf{2.71}                  \\
                               & Music 10dB  & 28.99                         & \textbf{9.92}                    & 26.16                      & \textbf{6.05}                  \\
                               & Music 5dB   & 34.77                         & \textbf{18.88}                   & 43.44                      & \textbf{12.37}                 \\
                               & Music 0dB   & 39.78                         & \textbf{33.57}                   & 50.4                       & \textbf{22.26}                 \\ \cmidrule(lr){2-2} \cmidrule(lr){3-4} \cmidrule(lr){5-6}
                               & Noise 20dB  & 13.79                         & \textbf{5.14}                    & 9.84                       & \textbf{1.28}                  \\
                               & Noise 15dB  & 16.99                         & \textbf{6.9}                     & 13.27                      & \textbf{2.3}                   \\
                               & Noise 10dB  & 21.16                         & \textbf{9.73}                    & 22.13                      & \textbf{3.59}                  \\
                               & Noise 5dB   & 25.67                         & \textbf{16.15}                   & 33.31                      & \textbf{6.98}                  \\
                               & Noise 0dB   & 30.31                         & \textbf{25.18}                   & 41.78                      & \textbf{12.12}                 \\ \cmidrule(lr){2-2} \cmidrule(lr){3-4} \cmidrule(lr){5-6}
                               & RT60 0.25 s & 8.52                          & \textbf{3.35}                    & 7.37                       & \textbf{3.76}                  \\
                               & RT60 0.5 s  & \textbf{11.89}                & 15.67                            & 50.69                      & \textbf{20.52}                 \\
                               & RT60 0.75 s & \textbf{17.36}                & 36.48                            & 73.83                      & \textbf{34.72}                 \\
                               & RT60 1 s    & \textbf{22.7}                 & 50.64                            & 76.76                      & \textbf{42.68}    \\ \bottomrule \bottomrule            
\end{tabular}
\end{table}

\subsubsection{Comparison of the Proposed Pre-trained Joint Optimization Method with Other Methods}
Table 6 shows the performance of the three systems used in this study on the noise test set under three training conditions: clean training data from ASVspoof 19 LA, data augmentation, and joint optimization with the speech enhancement front-end. We observe that for the ResNet18 and LCNN anti-spoofing networks without pre-training, noise data augmentation significantly improves model robustness. Further using the front-end speech separation U-net network can further enhance model robustness. However, we find that the proposed joint optimization model of the front-end speech enhancement and pre-trained Conformer shows less robustness improvement in many cases compared to simple data augmentation. On the reverberation test set, as shown in Table 7, we observe similar patterns, and the joint optimization model of the front-end speech enhancement and pre-trained Conformer also shows instability in robustness improvement under reverberation conditions.
\begin{table}[t]
\footnotesize
\centering
\caption{EER\% of }
\label{tab:5}
\begin{tabular}{lllllllllll}
\toprule \toprule
\multicolumn{2}{l}{\textbf{Model}}                     & \multicolumn{3}{c}{\textbf{ResNet18}}                                                  & \multicolumn{3}{c}{\textbf{LCNN}}                                                      & \multicolumn{3}{c}{\textbf{T. Conformer}}                                              \\ \cmidrule(lr){1-2} \cmidrule(lr){3-5} \cmidrule(lr){6-8} \cmidrule(lr){9-11}
\multicolumn{2}{l}{Data Augmentation}                       & \multicolumn{1}{c}{-} & \multicolumn{1}{c}{noise} & \multicolumn{1}{c}{noise} & \multicolumn{1}{c}{-} & \multicolumn{1}{c}{noise} & \multicolumn{1}{c}{noise} & \multicolumn{1}{c}{-} & \multicolumn{1}{c}{noise} & \multicolumn{1}{c}{noise} \\ \cmidrule(lr){1-2} \cmidrule(lr){3-5} \cmidrule(lr){6-8} \cmidrule(lr){9-11}
\multicolumn{2}{l}{SE}                        & \multicolumn{1}{c}{-} & \multicolumn{1}{c}{-}     & \multicolumn{1}{c}{Unet}  & \multicolumn{1}{c}{-} & \multicolumn{1}{c}{-}     & \multicolumn{1}{c}{Unet}  & \multicolumn{1}{c}{-} & \multicolumn{1}{c}{-}     & \multicolumn{1}{c}{Unet}  \\ \hline \hline
\multirow{16}{*}{test data set}
                                & Babble 20 dB & 15.98                 & 7.20                      & \textbf{6.21}             & 17.60                 & 7.31                      & \textbf{6.37}             & 5.48                  & \textbf{4.42}             & 5.12                      \\
                                & Babble 15 dB & 20.21                 & 9.08                      & \textbf{8.20}             & 23.54                 & 10.21                     & \textbf{9.69}             & 9.21                  & \textbf{6.05}             & 9.10                      \\
                                & Babble 10 dB & 25.40                 & 12.21                     & \textbf{11.38}            & 29.67                 & 13.79                     & \textbf{13.71}            & 15.24                 & \textbf{9.79}             & 15.57                     \\
                                & Babble 5 dB  & 30.64                 & 17.10                     & \textbf{16.46}            & 35.53                 & \textbf{18.81}            & 18.82                     & 24.35                 & \textbf{16.35}            & 23.15                     \\
                                & Babble 0 dB  & 35.48                 & 24.36                     & \textbf{23.73}            & 40.54                 & \textbf{24.70}            & 25.00                     & 34.19                 & \textbf{24.7}             & 29.82                     \\ \cmidrule(lr){2-2} \cmidrule(lr){3-5} \cmidrule(lr){6-8} \cmidrule(lr){9-11}
                                & Music 20 dB  & 25.72                 & 7.18                      & \textbf{5.75}             & 22.12                 & 6.46                      & \textbf{4.55}             & 4.58                  & 4.07                      & \textbf{3.39}                      \\
                                & Music 15 dB  & 33.13                 & 8.78                      & \textbf{6.66}             & 31.26                 & 8.25                      & \textbf{5.47}             & 6.48                  & 4.22                      & \textbf{3.76}                      \\
                                & Music 10 dB  & 40.05                 & 11.23                     & \textbf{8.29}             & 40.16                 & 11.14                     & \textbf{7.22}             & 9.92                  & \textbf{4.40}             & 4.49                      \\
                                & Music 5 dB   & 44.89                 & 15.27                     & \textbf{11.13}            & 45.58                 & 15.77                     & \textbf{10.13}            & 18.88                 & \textbf{5.29}             & 6.11                      \\
                                & Music 0 dB   & 48.80                 & 23.13                     & \textbf{16.71}            & 48.81                 & 23.78                     & \textbf{16.14}            & 33.57                 & \textbf{9.15}             & 10.28                     \\ \cmidrule(lr){2-2} \cmidrule(lr){3-5} \cmidrule(lr){6-8} \cmidrule(lr){9-11}
                                & Noise 20 dB  & 17.27                 & 6.76                      & \textbf{5.77}             & 15.53                 & 6.42                      & \textbf{4.83}             & 5.14                  & 4.20                      & \textbf{3.55}                      \\
                                & Noise 15 dB  & 22.19                 & 7.84                      & \textbf{6.68}             & 21.04                 & 7.82                      & \textbf{5.76}             & 6.90                  & 4.29                      & \textbf{3.88}                      \\
                                & Noise 10 dB  & 27.64                 & 9.64                      & \textbf{8.06}             & 27.02                 & 9.88                      & \textbf{7.05}             & 9.73                  & 4.56                      & \textbf{4.52}                      \\
                                & Noise 5 dB   & 32.75                 & 12.88                     & \textbf{10.30}            & 33.07                 & 13.34                     & \textbf{9.60}             & 16.15                 & \textbf{5.16}             & 5.63                      \\
                                & Noise 0 dB   & 37.6                  & 18.03                     & \textbf{14.50}            & 38.32                 & 18.57                     & \textbf{14.29}            & 25.18                 & \textbf{7.25}             & 7.99        \\             
\bottomrule \bottomrule
\end{tabular}
\end{table}

\begin{table}[t]
\footnotesize
\centering
\caption{EER\% of }
\label{tab:7}
\begin{tabular}{lllllllllll}
\toprule \toprule
\multicolumn{2}{l}{\textbf{Model}}                       & \multicolumn{3}{c}{\textbf{ResNet18}}    & \multicolumn{3}{c}{\textbf{LCNN}}        & \multicolumn{3}{c}{\textbf{T. Conformer}} \\ \cmidrule{1-2} \cmidrule(lr){3-5} \cmidrule(lr){6-8} \cmidrule(lr){9-11}
\multicolumn{2}{l}{Data Augmentation}                    & -     & reverb & reverb         & -     & reverb & reverb         & -      & reverb  & reverb        \\ \cmidrule{1-2} \cmidrule(lr){3-5} \cmidrule(lr){6-8} \cmidrule(lr){9-11}
\multicolumn{2}{l}{SE}                          & -     & -      & Unet           & -     & -      & Unet           & -      & -       & Unet          \\ \hline \hline
\multirow{5}{*}{test data set}
                               & RT60 0.25 s & 11    & 9.22   & \textbf{6.68}  & 9.75  & 7.79   & \textbf{6.94}  & \textbf{3.35}   & 6.05    & 4.75 \\
                               & RT60 0.5 s & 30.79 & 10.48  & \textbf{8.45}  & 28.55 & 9.02   & \textbf{8.45}  & 15.67  & 6.24    & \textbf{5.81} \\
                               & RT60 0.75 s & 39.48 & 11.62  & \textbf{9.69}  & 37.29 & 10.01  & \textbf{9.34}  & 36.48  & 7.2     & \textbf{6.83} \\
                               & RT60 1 s     & 42.42 & 12.35  & \textbf{10.59} & 41.44 & 10.62  & \textbf{10.17} & 50.64  & 8.42    & \textbf{7.85} \\ \bottomrule \bottomrule
\end{tabular}
\end{table}

We believe this is because the parameters of the Unet speech enhancement front-end are randomly initialized at the beginning of training, leading to poor reconstruction of the input signal's mel spectrogram by the Unet in the early stages of training. Since the Conformer is pre-trained, the expected input for the Conformer is a speech FBANK feature. The inconsistency between the embedding output by the Unet and the expected input domain of the Conformer increases the difficulty of training. We anticipate solving this issue by using a pre-trained Unet front-end model. As shown in Figure 9, this boxplot illustrates the average performance on various noise and reverberation conditions after jointly optimizing the Unet speech enhancement front-end with different epochs of pre-training with the T. Conformer on the ASVspoof19LA training and development sets. We observe that for Music and Environmental Noise, the average EER of the system decreases with the number of training epochs of the Unet enhancement network. Although the performance of the system on Babble noise and reverberation test data shows a trend with the increase in Unet training epochs, overall, using a pre-trained Unet front-end for joint optimization with the T. Conformer yields better results than directly using an untrained front-end enhancement network for joint optimization. This is because, after a certain number of pre-training epochs, the Unet's output embedding more closely matches the input feature distribution the T. Conformer is familiar with during training.

\subsubsection{Ablation Experiments}
Table 9 shows the results of the ablation experiments. When we freeze the parameters of the pre-trained Unet speech enhancement front-end, i.e., without joint training, speech enhancement and synthesized speech detection become independent tasks, which reduces the model's performance in complex scenarios to some extent. We observe that the proposed model with pre-trained joint optimization of the speech enhancement front-end and the detection back-end shows significant advantages under noise and reverberation conditions. However, for music noise and babble noise at extremely low SNRs of 0dB or 5dB, joint training may lead to slight performance degradation. This is because babble noise consists of real speech recordings, and at extremely low SNRs, overlapping speech segments make it difficult for the model to determine which segment needs to be judged for authenticity. Therefore, all methods perform poorly under extremely low SNR babble noise conditions, lacking comparability. Visualization of the output of the Unet on the fabricated sample LA\_E\_2178426.wav with 0dB music noise added shows that this is because the music noise presents clear tonal variations in the FBANK, which are distinctly different from the rhythm and pitch variations of speech. This allows the standalone speech enhancement front-end to accurately remove music noise without introducing additional artifacts. Hence, further joint training is unnecessary; an efficient anti-spoofing back-end model can effectively determine the authenticity of the speech. Freezing the Unet in the ResNet18 and LCNN back-end countermeasure models leads to consistent performance degradation across various noise and reverberation conditions.
\begin{figure*}
\label{}
  \centering
  \includegraphics[width=1\linewidth]{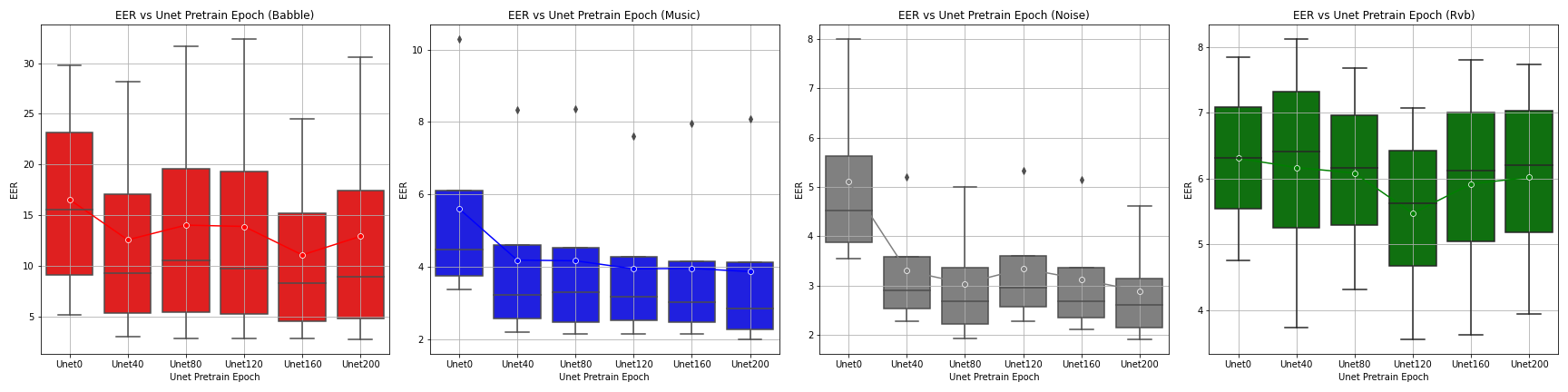}
  \caption{ablation}
  \label{fig:ablation}
\end{figure*}
When the Conformer model is not pre-trained, the robustness improvement brought by joint optimization still exists compared to clean training data and enhanced data training countermeasure models. However, the overall performance is poor due to the Conformer model's tendency to overfit on small-scale data without pre-training.

\subsubsection{Cross-Dataset Comparison}
To demonstrate the generalization ability of the proposed model, we conducted cross-dataset tests on two datasets: ASVspoof 2021 LA test set and FAD test set. This means the training and development datasets are ASVspoof 2019 LA, and only the testing is performed on the two out-of-domain datasets. As mentioned in Section 4.3, we use the ASVspoof 2021 LA test set to observe the performance changes of the proposed method when encountering non-additive noise such as channel coding. The FAD noise unseen test set is used to test the performance of the proposed method on cross-lingual and unknown noise types.

Table 10 shows the results on the ASVspoof 2021 LA test dataset. In addition to the models appearing in Section 5.3, two official best baselines (B03, B04) and the Wav2Vec2+AASIST model proposed by Tek et al. in 2021, which achieved the best single-model performance of 0.82 EER on the ASVspoof21LA test set, are listed. To compare cross-dataset experimental results, the results in Table 10 are without using the "Rowboost" data augmentation specifically designed for channel variations and compression coding. It is the SOTA single model on ASVspoof 2019. We use the open-source code of "AASIST" to conduct cross-dataset experiments on the ASVspoof 2021 LA test set. The results show that the baseline with speech enhancement has better generalization ability, and the proposed pre-trained joint optimization system achieves the lowest EER (3.9

\section{Conclusion}
This paper proposes a method for spoofed speech detection based on transfer learning of an ASR pre-trained Conformer model and joint optimization with a speech enhancement front-end to address the performance degradation of other anti-spoofing systems in noisy and reverberant scenarios. The transfer learning of the Conformer model backend brings more diverse and extensive out-of-domain positive sample data information to the model. The joint optimization of the speech enhancement front-end, based on our previous work, extends the discussion from additive noise to convolutional reverberation. The pre-trained Unet speech enhancement front-end addresses the input domain mismatch with the transfer learning back-end model. Experimental results show that the proposed method outperforms other baseline models in most SNR noise and different reverberation intensity scenarios and demonstrates cross-noise type generalization ability in cross-dataset experiments. However, the proposed method still needs improvement in handling babble noise. In the future, we will focus on addressing the challenge of detecting spoofed speech with real speech disturbances.



\bibliographystyle{elsarticle-num} 
\bibliography{my_ref.bib}

\end{document}